\documentstyle[12pt]{article}
\newcounter{saveeqn}
\newcommand{\alpheqn}{\setcounter{saveeqn}{\value{equation}}%
\setcounter{equation}{0}%
\addtocounter{saveeqn}{1}
\renewcommand{\theequation}{\mbox{\arabic{saveeqn}\alph{equation}}}}
\newcommand{\reseteqn}{\setcounter{equation}{\value{saveeqn}}%
\renewcommand{\theequation}{\arabic{equation}}}
\begin{document}
\thispagestyle{empty}
\vspace*{-1.5cm}
{\normalsize     
\hfill \parbox{50mm}{DESY 97-011\\January 1997}\\[15mm]
\begin{center} 
{\Large\bf The Lightest Neutralino in Supersymmetric \\
Standard Model with Arbitrary Higgs Sectors}
\\[14mm]
{\large P.N. Pandita}              
\\[5mm]
Deutsches Elektronen-Synchrotron DESY,\\
Notkestrasse 85, 22603 Hamburg,\\
Federal Republic of Germany\\
and\\
Department of Physics, North Eastern Hill University,\\
Umshing - Mawkynroh, Shillong 793 022, India\footnote{Permanent address} 
\\[15mm]
{\bf Abstract}
\end{center}

We consider the neutralino mass matrix in a supersymmetric 
theory based on
SU(2) ${\times}$ U(1) gauge group with arbitrary number of                  
singlet, doublet and triplet
Higgs superfields. We derive an upper bound on the 
mass of the lightest neutralino, and a lower bound on the mass of the heaviest
neutralino,  
in such a theory. Assuming grand unification of the 
gauge couplings, the upper bound on the mass of the lightest neutralino
can be expressed in terms of the gluino mass. For a gluino mass 
of 200 GeV, the
tree level upper bound on the mass of the lightest neutralino is 92 GeV, 
which increases to 166.5 GeV for a 1 TeV gluino. We also discuss the effect of
dominant one-loop
radiative corrections on these bounds.
\vfill
\newpage

\indent 
In supersymmetric gauge theories, each fermion and boson of the 
Standard Model is  
accompanied  by its supersymmetric partner, transforming in an  
identical manner under the gauge  group \cite{1}. In 
supersymmetric theories with R-parity conservation \cite{1}, 
the lightest 
supersymmetric particle (LSP) is expected to be the lightest
neutralino, which is the lightest mixture of the fermionic 
partners of the neutral Higgs and neutral electroweak gauge 
bosons. In order to give masses to quarks and
leptons, and to cancel triangle gauge anomalies, 
at least two Higgs
doublets $H_1=(H^0_1,H^-_1)$ and $H_2=(H^+_2,H^0_2)$, 
with opposite 
hypercharge $(Y(H_1)= -1, Y(H_2) = +1)$, are required 
in the minimal version of the Supersymmetric
Standard Model (MSSM) \cite{1}. The fermionic partners of these 
Higgs bosons mix with
the fermionic partners of the gauge bosons to produce 
four neutralino states
$\tilde{\chi}^0_i, ~i = 1, 2, 3, 4,$ and two chargino states
$\tilde{\chi}^{\pm}_i, ~i = 1, 2,$ in the MSSM. In the nonminimal 
supersymmetric standard model containing
a Higgs singlet, besides the two Higgs doublets of the 
minimal model, the
mixing of fermionic partners of neutral Higgs and gauge 
bosons produces 
five neutralino states. The neutralino states of the 
minimal model \cite{2,3,4} and
those of the nonminimal model \cite{5,6,7} have been studied 
in great detail,
because the lightest neutralino, being the LSP, 
is the end product of any
process involving supersymmetric particles in the final state.
\smallskip\\

Recently we considered the lightest neutralino state in a
general supersymmetric theory containing an arbitrary
number of singlet, doublet and triplet Higgs superfields
under the Standard Model gauge group, and obtained an upper 
bound on its mass \cite{8}. If we assume the simplest form of grand
unification of gauge couplings in such a theory, whereby the triplet 
and extra doublet Higgs fields are eliminated, then this upper bound is 
controlled by a soft supersymmetry gaugino mass (which we can take to be 
the gluino mass), and the vacuum expectation value of the doublet,
{\it but not the singlet}, Higgs fields. 
Since the latter are known, the bounds are controlled by the gluino mass 
alone. For a gluino mass of 200 GeV, the upper bound on the lightest neutralino
mass (including the dominant one-loop radiative corrections) was shown to be
62 GeV, which increases to 178 GeV for a  
1 TeV gluino. 
\smallskip\\
   
In this paper we generalize the above result. 
We consider the neutralino mass matrix in a 
general supersymmetric theory with an arbitrary Higgs sector, which
includes singlets, doublets and triplets under the 
Standard Model gauge group. We obtain an upper bound on 
the mass of the lightest neutralino and a lower 
bound on the mass of the heaviest neutralino state in such 
a general supersymmetric theory. These bounds depend on the 
soft supersymmetry breaking gaugino masses and the vacuum 
expectation values of the doublet and triplet, {\it but not 
the singlet},  Higgs fields. This is in contrast to the 
situation that obtains in the Higgs sector of such a general 
supersymmetric theory \cite{9}, where the (tree
level) upper bound on the lightest Higgs boson mass is 
controlled by the vacuum expectation value of the doublet 
Higgs fields and dimensionless parameters only. Nevertheless, 
if we assume grand unification of such a general
supersymmetric theory, then we show that the upper bound 
is controlled by a soft 
supersymmetry breaking gaugino mass parameter (which can be taken 
to be the gluino mass), and $M_Z$ and $\theta_W$. 
Since the latter are known, the former entirely determines the  upper
bound.
\smallskip\\

We start by recalling the neutralino mass matrix in the 
Minimal Supersymmetric Standard Model \cite{1}. In the basis
\begin{equation}
\psi^0_j=(-i\lambda^{\prime},-i\lambda^3,\psi_{H^1_1},\psi_{H^2_2}), \quad
j = \mbox{1, 2, 3, 4,}
\label{1}
\end{equation}
where $\lambda^{\prime}$ and $\lambda^3$ are the two-component 
gaugino states
corresponding to the U(1)$_Y$ and the third component 
of SU(2)$_L$ gauge groups, respectively, 
and $\psi_{H^1_1}$ and $\psi_{H^2_2}$ are the two-component
Higgsino states with hypercharge -1 and +1, respectively, 
the neutralino mass matrix can be written as
\smallskip
\begin{eqnarray}
\begin{footnotesize}
M = \left[ \begin{array}{cccc}
M_1 & 0 & -M_Z \sin\theta_W \cos \beta & M_Z \sin\theta_W \sin \beta \\
0 & M_2 & M_Z \cos \theta_W \cos \beta & -M_Z \cos\theta_W \sin \beta \\
-M_Z\sin\theta_W\cos\beta & M_Z\cos\theta_W\cos \beta & 0 & -\mu \\
M_Z\sin\theta_W\sin\beta &-M_Z\cos\theta_W\sin\beta & -\mu & 0\\
\end{array} \right]
\end{footnotesize}&&\nonumber\\[1ex]
&&\label{2}
\end{eqnarray}
\noindent
where $M_1$ and $M_2$ are supersymmetry breaking gaugino 
masses associated with the U(1)$_Y$ and SU(2)$_L$ subgroups 
of the standard model, respectively, and 
$\mu$ is the Higgs(ino) mixing parameter in the superpotential. 
Here, $\tan \beta = v_2/v_1$, where $v_1 = \langle H^0_1\rangle$ 
and $v_2 = \langle H^0_2\rangle$ are the vacuum expectation
values of the neutral components of the two Higgs doublets, 
and $M^2_Z = (g^2+g^{\prime2})(v^2_1+v^2_2)/2$, 
with $g$ and $g^{\prime}$ being the gauge couplings associated
with the SU(2) and U(1) factors of the standard model gauge 
group, respectively. 
Diagonalization of the mass matrix (\ref{2}) gives the physical 
neutralino masses and eigenstates 
\cite{10}. We shall denote these neutralino eigenstates 
by $\chi^0_1, \chi^0_2,
\chi^0_3$, and $\chi^0_4$, labelled in order of increasing mass. 
Since some
of the neutralino masses resulting from diagonalization 
of the mass matrix (\ref{2}) can be negative, it is convenient 
to consider the squared mass
matrix $M^{\dagger}M$ and the corresponding squared 
masses of the neutralinos.
An upper bound on the squared mass of the lightest 
neutralino $\chi^0_1$
can be obtained by using the fact that the smallest 
eigenvalue of $M^{\dagger}M$
is smaller than the smallest eigenvalue of its upper 
left 2 x 2 submatrix
\medskip\\
\begin{equation}
\left[ \begin{array}{cc}
M^2_1+M^2_Z\sin^2\theta_W & -M^2_Z\sin\theta_W\cos\theta_W \\
-M^2_Z\sin\theta_W\cos\theta_W & M^2_2 + M^2_Z\cos^2\theta_W
\end{array} \right]
\label{3}
\end{equation}
\medskip\\
thereby resulting in the upper bound \cite{11}
\\[0.5ex]
\begin{equation}
M^2_{\chi^0_1} \le \min(M^2_1+M^2_Z\sin^2\theta_W, M^2_2+M^2_Z\cos^2\theta_W).
\label{4}
\end{equation}
\\[0.5ex]
On the other hand, the larger eigenvalue of the 
upper left 2 x 2 submatrix
(\ref{3}) of $M^{\dagger}M$ gives a lower bound on 
the squared mass of the
heaviest neutralino:
\\[0.5ex]
\begin{equation}
M^2_{\chi^0_4} \ge \max(M^2_1+M^2_Z\sin^2\theta_W, M^2_2+M^2_Z\cos^2\theta_W).
\label{5}
\end{equation}
\\[0.5ex]
We note that the bounds (\ref{4}) and (\ref{5}) are controlled by, 
in addition
to $M_Z$ and $\theta_{W}$, the soft SUSY breaking gaugino masses, $M_1$ 
and $M_2$. 
This is in contrast
to the Higgs sector of MSSM, where the corresponding bounds 
on the (tree level)
masses of the lightest and the heaviest scalar Higgs bosons 
are controlled by
$M_Z$, and not by supersymmetry breaking masses \cite{12}.
\smallskip\\

We now consider a general class of supersymmetric standard models, namely, 
supersymmetric models based on Standard Model gauge group with an arbitrary
Higgs sector. We shall assume \cite{9}:
\begin{enumerate}
\item[(i)] Two Higgs doublets $H^{(1)}_1, H^{(1)}_2$, with hypercharge 
$Y=\mp1$, which are coupled to the quarks and leptons in the superpotential
\begin{equation}
W^0 = h_UQ_LU^c_LH^{(1)}_2 + h_DQ_LD^c_LH^{(1)}_1 + h_EL_LE^c_LH^{(1)}_1.
\label{6}
\end{equation}
In addition, we assume an arbitrary number of extra pairs of Higgs doublets
$H^{(j)}_1, H^{(j)}_2, j= 2,...,d+1$, which are decoupled from quarks and
leptons so that there are no dangerous flavor changing neutral currents
\cite{13}.
\item[(ii)] Gauge singlets $N^{(\sigma)}, \sigma = 1,...,n_s$.
\item[(iii)] SU(2) triplets $\Sigma^{(a)}, a = 1,...,t_0$, with $Y = 0$.
\item[(iv)] SU(2) triplets $\Psi^{(m)}_1, \Psi^{(m)}_2, m = 1,...,t_1$,
with $Y = \pm 2$.
\end{enumerate}
\mbox{ }\\[2ex]
Since the two Higgs doublets $H^{(1)}_1$ and $H^{(1)}_2$ are the minimum
number of doublets which are required in MSSM
to give masses to all fermions and to cancel triangle gauge anomalies,
the above extra Higgs multiplets are the only ones that can provide
renormalizable couplings to all possible combinations of  $H^{(1)}_1$ and
$H^{(1)}_2$, viz.  $H^{(1)}_1 H^{(1)}_2,  H^{(1)}_1 H^{(1)}_1$, 
and  $H^{(1)}_2H^{(1)}_2$,
in the superpotential. More exotic Higgs representations 
contribute to the $\beta$-functions of the gauge couplings 
and give lower values of the upper bound on the 
lightest Higgs mass in general supersymmetric theories. 
Since we are interested  in absolute upper bounds on the 
particle masses in these theories, we do not consider 
such representations. 
The most general renormalizable superpotential for the 
above Higgs supermultiplets
can be written as
\smallskip\\
\alpheqnŠ\begin{equation}
W^{\prime} = W_1 + W_2,
\label{7a}
\end{equation}
\\[0.5ex] 
where (repeated indices summed)
\smallskip\\
\begin{eqnarray}
W_1(H_{1,2},N,\Sigma,\Psi_{1,2}) &=& f^{ij\sigma}_1H^{(i)}_1H^{(j)}_2
N^{(\sigma)} + f^{ija}_2H^{(i)}_1\Sigma^{(a)}H^{(j)}_2 \nonumber \\
& & + g^{ijm}_1H^{(i)}_1\Psi^{(m)}_1H^{(j)}_1 + g^{ijm}_2H^{(i)}_2\Psi^{(m)}_2
H^{(j)}_2,
\label{7b}
\end{eqnarray}
\begin{eqnarray}
W_2(N,\Sigma,\Psi_{1,2}) &=& h_{alm}~tr(\Sigma^{(a)}\Psi^{(l)}_1\Psi^{(m)}_2)
+ \frac{\chi_{abc}}{6}~tr(\Sigma^{(a)}\Sigma^{(b)}\Sigma^{(c)}) \nonumber \\
& & + \frac{\lambda_{\mu\nu\sigma}}{6}N^{(\mu)}N^{(\nu)}N^{(\sigma)},
\label{7c}
\end{eqnarray}
\reseteqn 
\smallskip\\
and where we have represented the triplets by 2 x 2 traceless complex matrices,
and the trace in (\ref{7c}) is over the matrix indices. More explicitly, the
triplets are represented as (with multiplicity indices suppressed):
\alpheqn
\begin{equation}
\Sigma = \left[ \begin{array}{cc}
\xi^0/\sqrt{2} & \xi^+_2 \\
\xi^- & -\xi^0/\sqrt{2} \end{array} \right],
\label{8a}
\end{equation}
\bigskip
\begin{equation}
\Psi_1 = \left[ \begin{array}{cc}
\psi^+_1/\sqrt{2} & -\psi^{++}_1 \\
\psi^0_1 & -\psi^+_1/\sqrt{2} \end{array} \right],
\quad \Psi_2 = \left[ \begin{array}{cc}
\psi^+_2/\sqrt{2} & -\psi^{++}_2 \\
\psi^0_2 & -\psi^+_2/\sqrt{2} \end{array} \right].
\label{8b}
\end{equation}
\reseteqn 
\bigskip\\
Without loss of generality, we can make a unitary transformation in the
space of Higgs doublets $H^{(j)}_1$ and $H^{(j)}_2$ such that only the Higgs
doublets $H^{(1)}_1$ and $H^{(1)}_2$ aquire \cite{14} a non-zero vacuum 
expectation value ($\langle H^{(1)0}_1\rangle = v_1$, $\langle H^{(1)0}_2\rangle = v_2$). This requires \cite{15} that some of the Yukawa couplings
in (\ref{7b}) vanish:
\smallskip\\
\begin{equation}
f^{1j\sigma}_1 = f^{1ja}_2 = g^{1jm}_1 = g^{1jm}_2 = 0 \quad (j \neq 1).
\label{9}
\end{equation}
\smallskip\\
We will assume this condition in what follows. With this condition, the part
of the 
superpotential (\ref{7a}) involving only the neutral fields can be written as
\medskip\\
\begin{eqnarray}
W^{\prime} &=& -\sum_{\sigma}\left [\sum^{d+1}_{i = 1}f^{ii\sigma}_1H^{(i)0}_1H^{(i)0}_2 +
\sum_{i \ne 1, i \ne j}f^{ij\sigma}_1H^{(i)0}_1H^{(j)0}_2 \right ]N^{\sigma} 
\nonumber \\
& & + \sum_{a} \left [ \sum^{d+1}_{i = 1}f^{iia}_2 H^{(i)0}_1H^{(i)0}_2 + \sum_{i \ne 1, i \ne j}
f^{ija}_2 H^{(i)0}_1H^{(i)0}_2 + \right ]\frac{\xi^{(a)0}}{\sqrt{2}} 
\nonumber \\ 
& & - \sum_{m}\left [ \sum^{d+1}_{i = 1}g^{iim}_1H^{(i)0}_1H^{(i)0}_1 + 
\sum_{i \ne j \ne 1}g^{ijm}_1H^{(i)0}_1H^{(j)0}_2 \right ]\psi^{(m)0}_1 
\nonumber \\
& & - \sum_{m}\left [ \sum^{d+1}_{i = 1}g^{iim}_2H^{(i)0}_2H^{(i)0}_2 +
\sum_{i \ne j \ne 1}g^{ijm}_2H^{(i)0}_2H^{(j)0}_2 \right ] \psi^{(m)0}_2 
\nonumber \\
& & + \sum_{a}\left [\sum_{lm}h_{alm} \psi^{(l)0}_1
\psi^{(m)0}_2 \right ]\frac{\xi^{(a)0}}{\sqrt{2}} \nonumber \\ 
& & + \sum_{\mu \nu \sigma}\left [
\frac{\lambda_{\mu \nu \sigma}}{6}N^{(\mu)}
N^{(\nu)}N^{(\sigma)} \right ].        
\label{10}
\end{eqnarray}
Choosing the neutral fields
\medskip
\begin{eqnarray}
&&\psi^0_I~ =~ (-i\lambda^{\prime}, - i\lambda^3, \psi_{H^{(1)0}_1}, \psi_{H^{(j)0}_1}, 
\psi_{H^{(2)0}_2}, \psi_{H^{(j)0}_2},
\psi_{N^{(\sigma})}, \psi_{\xi^{(a)0}},\psi_{\psi^{(l)0}_1}, \psi_{\psi^{(l)0}_2}),
\nonumber\\[1ex]
&&j = 2,....,d+1,~~~\sigma = 1,....,n_s,~~~a = 1,....,t_0,~~~l = 1,...,t_1,~~~  
\\[1ex]
&&I~~ =~~ 1,......,(4+2d+n_s+t_0+2t_1),\mbox{     }\nonumber
\end{eqnarray}
as the basis, we can write the neutralino mass matrix for the general class
of supersymmetric models based on standard model gauge group with an
arbitrary Higgs sector as
\bigskip
\begin{eqnarray}
\begin{scriptsize}
M = \left[ \begin{array}{cccccccccc} 
M_1 & 0 & M_{13} & 0 & M_{14} & 0 & 0 & 0
& 2\sqrt{2}g^{\prime}y^l_1 & -2\sqrt{2}g^{\prime}y^l_2 \\[2ex]
& M_2 & M_{23} & 0 & M_{24} & 0 & 0 & g\sqrt{2}u^a &
-g\sqrt{2}y^i_1 & g\sqrt{2}y^i_2 \\[2ex]  
& & 2g^l_1y^l_1 & 0 & f & 0 &
f^{\sigma}_1v_2 & f^a_2v_2  & 2g^l_1v_1 & 0 \\[2ex]
& & & -2g^{ijl}_1y^l_1 & f^{i} & f^{ij} & f^{i\sigma}_1v_2 & 
f^{ia}_2v_2 & 0 & 0 \\[2ex] 
& & & & 2g^{l}_2y^l_2 & 0 & f^{\sigma}_1v_1 & f^a_2v_1 & 0 & 
2g^l_2v_2 \\[2ex]
& & \mbox{\scriptsize SYMM} & & & -2g^{ijl}_2y^l_2 & 0 & 0 & 0 & 0 \\[2ex]
& & & & & &  \lambda_{\sigma \gamma \delta}x^{\delta} & 0 & 0 & 0 \\[2ex]
& & & & & & & 0 & h^{\prime}_{alm}y^m_2 &
h^{\prime}_{aml}y^m_1 \\[2ex]
& & & & & & & & 0 & h^{\prime}_{aml}u^a \\[2ex]
& & & & & & & & & 0 
\end{array} \right]
\end{scriptsize}&&\nonumber\\[2ex]
&&\label{12}
\end{eqnarray}
where
\begin{center}
$M_{13}~~ =~~ -M_Z \sin \theta_W \cos \beta$,~~~~~$M_{14}~~ =~~  M_Z \sin
\theta_W 
\sin\beta$,
\\[2ex]
$M_{23}~~ =~~ M_Z \cos \theta_W \cos \beta$,~~~~~ $M_{24}~~ =~~ -M_Z \cos\theta_W\
sin\beta$,
\end{center}
\mbox{ }\\[2ex]
are the matrix elements which enter the neutralino mass matrix 
(\ref{2}) of the MSSM, and
we have defined the various vacuum expectation values and the couplings as
follows:
\alpheqn
\begin{equation}
u^a = \langle\xi^{(a)0}\rangle,~~  x^{\sigma} = \langle N^{(\sigma)}\rangle,~~ 
y^l_1 = \langle\psi^{(l)0}_1\rangle,~~ y^l_2 = \langle\psi^{(l)0}_2
\rangle,
\label{13a}
\end{equation}
\medskip
\begin{equation}
f^{\sigma}_1 = -f^{11\sigma}_1,~~ f^a_2 = f^{11a}_2/\sqrt{2},~~
f^{i\sigma}_1 = -f^{i1\sigma}_1,~~ 
f^{ia}_2 = f^{i1a}_2/\sqrt{2}, \:
\label{13b}
\end{equation}
\medskip
\begin{equation}
f = (f^{\sigma}_1x^{\sigma} +  f^a_2u^a),~~ 
f^{i} = (f^{i\sigma}_1x^{\sigma} +
f^{ia}_2u^a),~~  f^{ij} = (-f^{ij\sigma}x^{\sigma} + f^{ija}u^a/\sqrt{2}),~~ 
\end{equation}
\label{13c}
\medskip
\begin{equation}
g^l_1 = -g^{11l}_1,~~ \:g^l_2 = -g^{11l}_2,~~ \: h^{\prime}_{alm} = h_{alm}
/\sqrt{2}. \: 
\label{13d}
\end{equation}
\reseteqn
\medskip\\
\noindent
In (\ref{13a}) we have represented the Higgs components of the 
superfields by the same symbol as the superfields themselves. The upper bound 
on the squared mass of the lightest neutralino can be obtained by examining 
the upper left 2 x 2 submatrix of $M^{\dagger}M$ corresponding to $M$ of
Š(\ref{12}). This submatrix can be written as
\medskip\\
\alpheqn 
\begin{equation}
\left[ \begin{array}{cc}
M^2_1 + M^2_Z \sin^2 \theta_W + 6g^{\prime2}y^2 & 
-M^2_Z \sin \theta_W \cos \theta_W - 2gg^{\prime}y^2 \\[2ex]
-M^2_Z \sin \theta_W \cos \theta_W - 2gg^{\prime}y^2 &
M^2_2 + M^2_Z \cos^2 \theta_W + 2g^2u^2
\end{array} \right],
\label{14a}
\end{equation}
\medskip\\
where
\smallskip\\
\begin{equation}
y^2 = \sum_l [(y^l_1)^2 + (y^l_2)^2], \quad u^2 = \sum_a 
(u^a)^2,
\label{14b}
\end{equation}
\reseteqn
\medskip\\
and where we have used the expressions for $W$ and $Z$ masses appropriate
for the general theory that we are considering:
\alpheqn
\begin{equation}
M^2_Z = \frac12 (g^2 + g^{\prime2}) [v^2_1 + v^2_2 + 4y^2],
\label{15a}
\end{equation}
\begin{equation}
M^2_W = \frac12 g^2[v^2_1 + v^2_2 + 4u^2 + 2 y^2].
\label{15b}
\end{equation}
\reseteqn
\medskip\\
The vaccum expectation values that enter into the $W$ and $Z$ masses (\ref{15a})
and (\ref{15b}) are experimentally constrained by the $\rho$ parameter. From
a recent global fit, which includes the CDF data, we have \cite{16}
\begin{equation}
\rho = \frac{M^2_W}{M^2_Z\cos^2\theta_W} = 1.0002 \pm 0.0013 \pm 0.0018,
\label{16}
\end{equation}
where the second error is due to the Higgs mass. This result is remarkably
close to the expected Standard Model value of $\rho = 1$. Taking a value of
$\rho = 1$ implies, through (\ref{15a}) and (\ref{15b}), the following 
relation between the triplet vacuum expectation values:
\begin{equation}
4u^2 = 2y^2.
\label{17}
\end{equation}
We shall henceforth assume (\ref{17}) to be true.
With constraint (\ref{17}), the $W$ and
$Z$ masses can be written as \cite{17}
\[
M^2_W = M^2_Z \cos^2 \theta_W = \frac12 g^2(v^2_1 + v^2_2 + 4y^2).
\]
Thus, the combination $(v^2_1 + v^2_2 + 4y^2)$ is determined to be
$\simeq (174 GeV)^2$, but the ratio of the triplet to the doublet vacuum
expectation values $y/(v^2_1 + v^2_2)^{1/2}$ is not fixed. Using Š(\ref{17}),
the 2 x 2 submatrix (\ref{14a}) can be written as
\medskip\\
\begin{equation} 
\left[ \begin{array}{cc}
M^2_1 + M^2_Z \sin^2 \theta_W + 6g^{\prime2}y^2 & 
-M^2_Z \cos \theta_W \sin \theta_W - 2gg^{\prime}y^2 \\[2ex]
-M^2_Z \cos \theta_W \sin \theta_W - 2gg^{\prime}y^2 &
M^2_2 + M^2_Z \cos^2 \theta_W + g^2y^2
\end{array} \right].
\label{18}
\end{equation} 
\medskip\\
The smallest eigenvalue of the 2 x 2 matrix (\ref{18}) serves as the upper 
bound on the squared mass of the lightest neutralino in the general supersymmetric
Standard Model:
\begin{equation}
M^2_{\chi^0_1} \le \min(M^2_1 + M^2_Z \sin^2\theta_W + 6g^{\prime2}y^2,
M^2_2 + M^2_Z \cos^2 \theta_W + g^2y^2).
\label{19}
\end{equation}
On the other hand, the heaviest neutralino $(\chi^0_n)$ mass is bounded from
below by
\begin{equation}
M^2_{\chi^0_n} \ge \max(M^2_1 + M^2_Z \sin^2\theta_W + 6g^{\prime2}y^2,
M^2_2 + M^2_Z \cos^2 \theta_W + g^2y^2).
\label{20}
\end{equation}
These bounds depend on a priori unknown vacuum expectation values of the
triplet Higgs fields $(\psi^{(i)}_1, \psi^{(i)}_2)$,  and the supersymmetry
breaking gaugino 
mass parameters.
Nevertheless, as we shall see these bounds can become meaningful in theories
with gauge coupling unification \cite{18}. We note that the singlet vacuum
expectation values decouple from these bounds.
\smallskip\\

In the general supersymmetric theory that we are considering, 
the renormalization
group equations (RGEs) \cite{15} for the standard SU(3) x SU(2) x U(1) gauge
couplings can be written as $(g^2_1 = (5/3)g^{\prime2},~ g^2_2 = g^2,~ \tan
\theta_W = g^{\prime}/g,~ g_3$ is the SU(3) gauge coupling constant)
\begin{eqnarray}
16 \pi^2 \frac{dg_1}{dt} = \left[ \frac{33}{5} + \frac35 (6t_1 + d)\right]
g^3_1, \nonumber \\
16 \pi^2 \frac{dg_2}{dt} = \left[ 1 + 2t_0 + 4t_1 +d)\right]
g^3_2, \label{21} \\
16 \pi^2 \frac{dg_3}{dt} = -3g^2_3. \nonumber
\end{eqnarray}
Obviously, these  RGEs depend on the number and the type of 
the Higgs representations
$(d, t_0, t_1)$. We note that the additional doublets increase the $\beta$
functions of the gauge couplings, even though the VEVs of the additional
doublets have been rotated away. If we assume that the gauge couplings 
$g_1, g_2$, and $g_3$ unify at some grand unification scale, i.e. $g_1(M_U)
= g_2(M_U) = f_3(M_U)$, with $M_U$ the unification scale, 
then the simplest choice is \cite{8}
\begin{equation}
d = t_0 = t_1 = 0,
\label{22}
\end{equation}
with $n_s$ arbitrary, i.e. MSSM with an arbitrary number of singlet superfields.
In other words, only Higgs singlets, besides the two Higgs doublets of MSSM,
are consistent with the simplest form of unification \cite{19}.ŠThe bounds (\ref{19}) and (\ref{20}) now reduce to
\begin{equation}
M^2_{\chi^0_1} \le \min(M^2_1 + M^2_2 \sin^2 \theta_W, \: M^2_2 + M^2_Z
\cos^2 \theta_W),
\label{23}
\end{equation}
\begin{equation}
M^2_{\chi^0_n} \ge \max(M^2_1 + M^2_2 \sin^2 \theta_W, \: M^2_2 + M^2_Z
\cos^2 \theta_W),
\label{24}
\end{equation}
which are identical to the corresponding bounds (\ref{4}) and (\ref{5}) in the
minimal supersymmetric standard model \cite{8}. 
\smallskip

However, gauge coupling unification can be achieved even when 
$d,  t_0,  t_1  >  0$
by adding colored multiplets or introducing an intermediate scale in the 
theory \cite{20}. Therefore, it is important to ask what are the bounds on
the mass of the lightest neutralino in the most general case without 
assuming any particular form of grand unification with a specific 
particle content. To evaluate the 
upper bound (\ref{19}) in the most general case, we recall that
the gaugino mass parameters satisfy the renormalization group equations (RGEs)
\cite{21}:
\begin{equation}
16 \pi^2 \frac{dM_i}{dt} = b_iM_ig^2_i,
\label{25}
\end{equation}
where the coefficients $b_i$ are the same that occur in the evolution 
(\ref{21}) of 
the corresponding gauge couplings. Equations (\ref{21}) and 
(\ref{25}) imply $(\alpha_i = g^2_i/4 \pi,
\alpha_U = g^2_U/4\pi)$,
\begin{equation}
M_1(M_Z)/\alpha_1(M_Z) = M_2(M_Z)/\alpha_2(M_Z) = M_3(M_Z)/\alpha_3(M_Z) =
m_{1/2}/\alpha_U,
\label{26}
\end{equation}
where $M_{1/2}$ is the common gaugino mass at the grand unification scale,
and $\alpha_U$ is the unified coupling constant. It is important to note
that (\ref{26}), which is a consequence of one-loop renormalization group 
equations, is valid in any grand unified theory irrespective of the 
particle content.
Two-loop effects to (\ref{26}) are expected to
be numerically
small \cite{22}. We also note that the GUT relation can be violated only
at order $\alpha_U/\pi$ \cite{23}. The relation (\ref{26})
is the same which occurs in the MSSM with grand unification. It
reduces the three gaugino mass parameters to one, and since the
the gluino mass $m_{\tilde{g}}$
is equal to $|M_3|$, it is, therefore, convenient to let the gluino mass be the
free parameter. The other two gaugino mass parameters are then determined
through
\begin{eqnarray}
M_2(M_Z) = (\alpha/\alpha_3 \sin^2 \theta_W)m_{\tilde{g}} \simeq 0.27m_{\tilde{g}},
\nonumber \\
M_1(M_Z) = (5\alpha/3\alpha_3 \cos^2 \theta_W)m_{\tilde{g}} \simeq 0.14m_{\tilde{g}},
\label{27}
\end{eqnarray}
where $\alpha$ is the fine structure constant, and where we have used
\cite{24,25,26,27} the values of couplings at the $Z^0$ energy as
\begin{equation}
\alpha^{-1}(M_Z) = 127.9, \: \sin^2\theta_W = 0.2315, \: \alpha_3 = 0.123.
\label{28}
\end{equation}
Using (\ref{27}) and the most conservative estimate of the upper bound 
on the triplet vacuum
expectation value, $y^2 \le (87 GeV)^2$, following from $(v^2_1 + v^2_2 + 
4 y^2) \simeq (174 GeV)^2$, the upper bound (\ref{19})  
can be written  as 

\begin{equation}
M^2_{\chi^0_1} \le M^2_1 + M^2_Z \sin^2 \theta_W + 6 g^{\prime 2}y^2 
\simeq (0.02 m^2_{\tilde{g}} +
7733.5)\mbox{GeV}^2.
\label{29}
\end{equation}
Note that the lower bound (\ref{20}) on the mass of the heaviest neutralino
cannot be evaluated in the most 
general case that we are considering. For a gluino mass of 200 GeV, 
the upper bound (\ref{29}) on the mass of
the  lightest
neutralino is 92 GeV. Similarly, for a gluino mass of 1 TeV, the upper bound 
becomes 166.5 GeV. This should be compared with the upper bound of 52 GeV
and 148 GeV, respectively, for gluino mass of 200 GeV and 1 TeV, which one 
obtains when one assumes \cite{8} a simple form of gauge coupling unification
(\ref{22}).
\smallskip\\

The bounds (\ref{19}), (\ref{20}) and (\ref{29}) are our main results.
While the former are valid in a supersymmetric 
theory based on a standard model gauge group and an arbitrary Higgs sector, 
the latter is valid only when we impose the constraint of grand unification 
on gauge couplings in such a supersymmetric theory. Thus, although
a bound on the lightest neutralino mass exists in an arbitrary supersymmetric
theory, it is calculable in terms of gluino mass only under the assumption
of gauge coupling unification. We further note that since the soft 
supersymmetry breaking gaugino masses satisfy the GUT relation (\ref{26})
in any grand unified theory based on a simple group independent of the
breaking pattern to the Standard Model gauge group \cite{28}, our result
(\ref{29}) is valid in any such supersymmetric grand unified
theory.
\smallskip\\

Finally, we discuss the effect of radiative corrections on the tree level
upper bound on the lightest neutralino mass derived
above. The dominant radiative corrections arise from top-stop loops. If
the neutralino is predominantly Higgsino-like, then a simple estimate of the
radiative corrections to its mass arising from a top-stop loop is given by

\begin{equation}
\frac{\Delta M_{\chi^0}}{M_{\chi^0}} \simeq 3 \frac{h^2_t}{16 \pi^2} \ln
\left( \frac{m^2_{\tilde{t}}}{m^2_t} \right) \simeq 5 \%,
\label{31}
\end{equation}
where the factor 3 comes from color, $h^2_t$ comes from the top Yukawa
couplings at the two vertices in the loop $(h_t \equiv h_{U33}$, see (\ref{6})),
and $1/16 \pi^2$ is the usual loop factor. In making the above estimate, we 
have taken the squark mass to be equal to 1 TeV. A similar estimate holds
for a gaugino-like neutralino, and also yields an estimate for the correction 
of order 5$\%$. These estimates are very close to the generic results which
emerge from detailed calculations \cite{29} carried out in the context of
the minimal supersymmetric standard model, although, for some extreme values 
of parameters, the radiative corrections to the lightest neutralino mass can
be as large as 20$\%$. 
Taking these results as indicative of the radiative corrections
in the general model that we are considering, we estimate a conservative
radiatively corrected upper bound on the mass of the lightest neutralino to be
about 110 GeV for a 200 GeV gluino, which increases to 200 GeV 
for a gluino
mass of 1 TeV \cite{30}.

\section*{Acknowledgements}
I would like to thank the theory group at DESY for hospitality while this work
was completed. 
This work is supported by the Department of Atomic
Energy under Grant No. 37/14/95-R \& D-II/663.


\newpage

\begin{thebibliography}{99}

\bibitem{1} H.P. Nilles, Phys. Ref. {\bf 110}, 1 (1984); H.E. Haber and G.L.
Kane, Phys. Ref. {\bf 117}, 75 (1985).
\bibitem{2} J. F. Gunion and H.E. Haber, Nucl. Phys. {\bf B272}, 1 (1986);
Nucl. Phys. {\bf B278}, 449 (1986); Phys. Rev. {\bf D37}, 2515 (1988).
\bibitem{3} A. Bartl, H. Fraas and W. Majerotto, Nucl. Phys. {\bf B270},
1 (1986); A. Bartl et. al., Phys. Rev. {\bf D40}, 1594 (1989).
\bibitem{4} M.M. Elkheisen, A.A. Shafik and A.A. Aboshonsha, Phys. Rev.
{\bf D45}, 4345 (1992); M. Guchait, Z. Phys. {\bf C57}, 158 (1993).
\bibitem{5} P.N. Pandita, Phys. Rev. {\bf D50}, 571 (1994).
\bibitem{6} P.N. Pandita, Z. Phys. {\bf C63}, 659 (1994).
\bibitem{7} F. Franke and F. Fraas, W\"urzburg Preprint, 
WUE-ITP-95-024 (1995).
\bibitem{8} P.N. Pandita, Phys. Rev. {\bf D53}, 566 (1996).
\bibitem{9} J.R. Espinosa and M. Quiros, Phys. Lett. {\bf B301}, 51 (1993);
G.L. Kane, C. Kolda and J.D. Wells, Phys. Rev. Lett. {\bf 70}, 2686 (1993).
\bibitem{10} We are assuing CP invariance in the neutralino sector such that
the neutralino mass matrix is a real symmetric matrix.
\bibitem{11} See also, R. Barbieri and G.F. Guidice, Nucl. Phys. {\bf B306},
63 (1988).
\bibitem{12} For a review and references, see, P.N. Pandita, Pramana, J. Phys.
Suppl. {\bf 41}, 303 (1993); H.E. Haber, in: {\it Perspective on Higgs Physics}
, ed. G.L. Kane (World Scientific, Singapore, 1992).
\bibitem{13} S.L. Glashow and S. Weinberg, Phys. Rev. {\bf D115}, 1958
(1977).
\bibitem{14} H. Georgi and D. Nanopoulos, Phys. Lett. {\bf B82}, 95 
(1979).
\bibitem{15} J.R. Espinosa and M. Quiros,  Ref. \cite{9}.
\bibitem{16} Particle Data Group, Review of Particle Properties, Phys. Rev.
{\bf D54}, 1 (1996).
\bibitem{17} See, H. Georgi and M. Machacek, Nucl. Phys. {\bf B262}, 463
(1995) in the context of non-supersymmetric Standard Model.
\bibitem{18} J. Ellis, S. Kelly and D.V. Nanopoulos, Phys. Lett. {\bf B260},
131 (1991); U. Amaldi, W. de Boer and H. F\"urstenau, Phys. Lett. {\bf B260},
447 (1991); P. Langacker and M. Luo, Phys. Rev. {\bf D44}, 817 (1991).
\bibitem{19} See, e.g., P. Langacker and N. Polonsky, Phys. Rev. {\bf D50},
2199 (1994); R. Arnowitt and P. Nath, Lectures at VII J. A. Sweica Summer 
School, Campos de Jordao, Brazil, 1993; CTP--TAMU--52/93 and NUB--TH--3073--93.
\bibitem{20} U. Amaldi et al., Phys. Lett. {\bf B281}, 374 (1992);
For a review and references, see, e.g., W. de Boer, Prog. Part. Nucl. Phys.
{\bf 33}, 201 (1994).
\bibitem{21} K. Inoue, A. Kakuto, H. Komatsu and S. Takeshita, Prog. Theor.
Phys. {\bf 68}, 927 (1982); ibid {\bf 71}, 413 (1984).
\bibitem{22} Y. Yamada, Phys. Lett. {\bf B316}, 109 (1993); Phys. Rev. Lett. 
{\bf 72}, 25 (1994); S.P. Martin and M.T. Vaughn, Phys. Lett. {\bf B318},
331 (1993); Phys. Rev. {\bf D50}, 2282 (1994).
\bibitem{23} J. Hisano, H. Murayama and T. Goto, Phys. Rev. {\bf D49}, 
1446 (1993).
\bibitem{24} G. Degrassi, S. Fanchiotti and A. Sirlin, Nucl. Phys. {\bf B352},
49 (1991).
\bibitem{25} P. Langacker and N. Polonsky, Phys. Rev. {\bf D47}, 4028 (1993).
\bibitem{26} The LEP Collaborations: ALEPH, DELPHI, L3 and OPAL, Phys. Lett.
{\bf B270}, 247 (1992); CERN/PPE/93--157.
\bibitem{27} Particle Data Group, Review of Particle Properties, ref. \cite{16}
\bibitem{28} Y. Kawamura, H. Murayama and M. Yamaguchi, Phys. Lett.
{\bf B324}, 52 (1994).
\bibitem{29} A.B. Lahanas, K. Tamvakis and N.D. Tracas, Phys. Lett. 
{\bf B324}, 387 (1994); D. Pierce and A. Papadopoulos, Phys. Rev.
{\bf D50}, 565 (1994), Nucl. Phys. {\bf B430}, 278 (1994).
\bibitem{30} The input gluino mass used here will be the radiatively corrected
physical mass of the gluino.
\end{thebibliography}
\end{document}